\documentstyle[pra,aps,psfig]{revtex}
\tighten

\newcommand{\ket}[1]{\left | #1 \right \rangle}
\newcommand{\bra}[1]{\left \langle #1 \right |}
\newcommand{\amp}[2]{\left \langle #1 | #2 \right \rangle}
\newcommand{\proj}[1]{\ket{#1} \bra{#1}}
\newcommand{\tr}{{\rm \, Tr }\, }
\begin{document}
\draft
\title{Quantum template matching}
\author{Masahide Sasaki$^{1, 2}$, Alberto Carlini$^{1, 2}$
and Richard Jozsa$^3$}
\address{$^1$Communications Research Laboratory,
Ministry of Posts and Telecommunications,\\
 Koganei, Tokyo 184-8795, Japan.\\
$^2$CREST, Japan Science and Technology\\
E-mail:psasaki@crl.go.jp\\[2mm]
$^3$Department of Computer
Science, University of Bristol,\\ Woodland Road, Bristol BS8 1UB,
England. }


\maketitle

\begin{abstract}
We consider the quantum analogue of the pattern matching problem,
which consists of classifying a given unknown system according to
certain predefined pattern classes. We address the problem of
quantum template matching in which each pattern class ${\cal C}_i$
is represented by a known quantum state $\hat g_i$ called a
template state, and our task is to find a template which optimally
matches a given unknown quantum state $\hat f$. We set up a
precise formulation of this problem in terms of the optimal
strategy for an associated quantum Bayesian inference problem. We
then investigate various examples of quantum template matching for
qubit systems, considering the effect of allowing a finite number
of copies of the input state $\hat f$. We compare quantum optimal
matching strategies and semiclassical strategies and demonstrate
an entanglement assisted enhancement of performance in the general
quantum optimal strategy.
\end{abstract}



\section{Introduction}\label{introduction}

Let us consider the following {\it pattern matching} problem. We
have at our disposal a database of recorded persons' pictures,
organized into different classes according to certain defined
features. Now we are given a further person's picture and we want
to determine to which class the person belongs.  We scan the
database by comparing the defined features of the given sample
with those of the classes. If the patterns of the sample have a
good enough matching with those of a certain class, then we can
say that the person is {\it recognized}, that is, that the person
belongs to that pattern class.  We would like to consider a
similar problem in the quantum mechanical context.

There are various kinds of approaches to this problem
\cite{Fu76,Fukunaga90}.  Given a pattern we wish to classify it
relative to a predefined set of pattern classes in such a way that we
maximize some suitable measure of matching. This could be formulated,
e.g., in terms of vector representations, that is, a given image is
discretized on a mesh and the contents in each pixel are approximated
by the value of some predetermined intensity levels, say $f(i,j)$ for
the $(i,j)$-th pixel of a two dimensional mesh. Thus a pattern is
represented by a vector $\vec f=(f(1,1), f(1,2), ...)^T$.  When an
input sample $\vec f_0$ is given, an intrinsic feature is first
extracted from it by removing noise and adjusting its size.  The
resulting data, say $\vec f$, called the {\it feature vector}, should
be less noisy, less redundant and more invariant under commonly
encountered variations and distortions.  We want then to classify
this $\vec f$ into a pattern class from a set of classes $\{{\cal
C}_i\}$, where each class ${\cal C}_i$ contains some similar
patterns. Classification is usually made by evaluating a {\it
discriminant function} $D_i(\vec f)$ associated with the pattern
class ${\cal C}_i$, and is such that if the input sample is actually
in the class $j$, the value $D_j(\vec f)$ must be the largest.  One
way to make the problem more tractable is to represent each class
${\cal C}_i$ by a typical pattern called a {\it template vector}
$\vec g_i$ and to deal only with this vector as the representative of
the class ${\cal C}_i$.  This is the template matching problem.

Now consider a similar problem in the quantum mechanical context.
We are given a feature state $\ket f$, which is usually unknown,
and a set of classes $\{{\cal C}_i\}$ and their associated
template states $\ket{g_i}$, known {\it a priori}.  The problem of
{\it quantum template matching} is to classify the state $\ket f$
according to the set $\{\ket{g_i}\}$, that is, to pick up the
template which best matches with $\ket f$.  This is similar in
some respects to quantum state discrimination and quantum state
estimation. In quantum state discrimination, a discrete set of
states $\{\hat\rho_i\}$ and their {\it a priori} probabilities
$\{p_i\}$ are given.  The task is to decide which state is
received.  In quantum state estimation, one is to reconstruct a
given unknown state $\hat\rho$ by estimating certain parameters.
In both scenarios, one usually minimizes a certain Bayes cost such
as a decision or estimation error by using prior knowledge about
the states.  In quantum template matching, although we deal with
an unknown input state generally specified by continuous
parameters, the purpose is not to estimate the input state or to
discriminate among the input states themselves, but to assign the
best matched template state from amongst given candidates. In this
sense quantum template matching involves aspects of both state
estimation and discrimination: the unknown input states are
generally parameterized by continuous parameters which we wish to
characterize only up to some approximation given by the
``closest'' template state. Indeed direct state estimation or
discrimination would provide a strategy for template matching (by
comparing the classical information of the estimated state with
the classical information of the template identities) but this is
generally not optimal -- we should attempt to best match a
template without necessarily obtaining any further more detailed
information about the identity of the input state itself.

In this paper we will set up a precise formulation of this problem
(in section \ref{Bayesian}) in terms of a suitable intuitive
matching criterion. The template matching problem will then appear
as a problem of determining the optimal strategy for an associated
quantum Bayesian inference problem
\cite{Helstrom_QDET,Holevo_book}. We will then consider some
examples of template matching for qubit systems (in sections
\ref{Binary_temp_match} and \ref{Multiple_temp_match}) in
particular, considering the effect of allowing a finite number of
identical copies of the input state $\ket{f}$ (of course in a
classical context this makes no difference). We will compare the
optimal strategy (allowing full use of entanglement across the
space of all copies) with two semiclassical strategies: \\ (a)
applying only separate measurements on each copy and processing
the outcomes to decide the best matching; \\ (b) applying the
optimal state estimation strategy
\cite{Massar95,Derka98,Latorre98} (using a collective measurement
on the product state of all copies) and then classically comparing
the identity of the reconstructed state with that of the template
states. \\ We will see that the optimal (fully entangled) strategy
is more efficient than either of these. Finally in section
\ref{Concluding} we will summarize our results and describe some
interesting further possible generalizations of the concept of
quantum template matching.

\section{Bayesian formulation of template matching}\label{Bayesian}

The Bayesian formulation is based on an {\it a priori} knowledge
about the inputs: the input feature state $\hat f$ is unknown but
it is assumed that we know the {\it a priori} probability
distribution $P(\hat f)$ for possible inputs. Each template state
$\hat g_i$ representing the class ${\cal C}_i$ is assumed to be
completely known. In order to classify $\hat f$ into a class
${\cal C}_j$, we need to introduce a {\it score} $S({\cal
C}_j\vert\hat f)$ which provides a matching criterion. One
reasonable choice for the score in template matching is that
derived from the similarity criterion

\begin{equation}
S({\cal C}_j\vert\hat f)\equiv \left( \tr{\left ( \sqrt{\sqrt{\hat
f}\hat g_j\sqrt{\hat f}}\right )} \right)^2 , \label{cond_score}
\end{equation}
which is just the fidelity between the input state $\hat f$  and
the template state $\hat g_i$. If the template states are pure
states $\ket{g_j}$ then this is just the standard overlap
$\bra{g_j} \hat{f} \ket{g_j}$. Under the similarity criterion, we
are to choose the template for which the state overlap with $\hat
f$ is largest. The matching strategy is represented by a
probability operator measure (POM) $\{\hat \Pi_j\}$:
\begin{equation}
\hat \Pi_j=\hat \Pi_j^\dagger\ge0, \quad
\sum_j \hat \Pi_j = \hat I.
\end{equation}
This should be designed by using the {\it a priori} knowledge on
the input and the template states, and the conditional scores
$S({\cal C}_j\vert\hat f)$.  The performance of a matching
strategy is measured by the {\it average score} defined as
\begin{equation}
\bar S\equiv \sum_j\int df S({\cal C}_j\vert\hat f) P({\cal
C}_j\vert\hat f)P(\hat f), \label {def_average_score}
\end{equation}
where $P({\cal C}_j\vert\hat f)\equiv\tr{(\hat \Pi_j\hat f)}$ is
the conditional probability that we have the $j$-th outcome given
the state $\hat f$. The best strategy is the one that maximizes
this average score. If we  introduce the score operators
\begin{equation}
\hat W_j\equiv\int df S({\cal C}_j\vert\hat f)P(\hat f)\hat f,
\label {score_op}
\end{equation}
then Eq.  (\ref{def_average_score} ) can be rewritten as
\begin{equation}
\bar S=\sum_j\tr{(\hat W_j\hat \Pi_j)}. \label {average_cost_Wj}
\end{equation}

Thus the problem is to find the optimal POM $\{\hat \Pi_j\}$ that
maximizes $\bar S$ given the set of score operators $\{\hat
W_j\}$. This is a standard quantum  Bayesian optimization problem
and necessary and sufficient conditions for optimality are well
known \cite{Holevo73_condition,Yuen,Helstrom_QDET}:
\begin{equation}
\begin{array}{cl}
\mbox{(i)}&\hat\Gamma\equiv\displaystyle\sum_j\hat W_j\hat \Pi_j
\mbox{ is hermitian},
\\ {}&{}\\ \mbox{(ii)}&\hat\Gamma-\hat W_j \quad
\ge0\quad\forall j.
\end{array}
\label{opt_cond}
\end{equation}
Note that since $S({\cal C}_j\vert\hat f)$ is always non-negative, we
have $\hat{W}_j \geq 0$ as operators. Hence our optimal template
matching problem (of optimizing eq. (\ref{average_cost_Wj})) reduces
to a standard quantum state discrimination problem -- of
distinguishing the mixed states $\hat{W}_j /\tr \hat{W}_j$ (the
normalized score operators) taken with prior probabilities $p_j = \tr
\hat{W}_j /\sum_j \tr \hat{W}_j $. In general the optimal strategy is
unknown but we will consider examples exhibiting symmetry in which
optimal strategies can be given.

\section{Binary template matching of a two state system}
\label{Binary_temp_match}

We begin with the simplest case of quantum template matching in
which there are only two classes, and each class is described by a
template state which is a known pure qubit state. Furthermore the
input states $\ket{f}$ will be restricted to depend on only a
single real feature parameter. By taking an appropriate qubit
basis $\{\ket{\uparrow},\ket{\downarrow}\}$, the binary template
states can be represented in terms of real components
\begin{mathletters}
\begin{eqnarray}
\ket{g_0}&=&{\rm cos}{\theta\over2}\ket{\uparrow}
                     +{\rm sin}{\theta\over2}\ket{\downarrow}, \\
\ket{g_1}&=&{\rm sin}{\theta\over2}\ket{\uparrow}
                     +{\rm cos}{\theta\over2}\ket{\downarrow},
\end{eqnarray}
\label{templates}
\end{mathletters}

\noindent with a single real parameter $\theta$ specifying the
nonorthogonality between the templates. As for the input state
$\ket{f}$, we will assume that its input distribution is the uniform
probability density over the great circle on the Bloch sphere defined
by the two template states. Thus we can write
\begin{equation}
\ket{f(\phi)}={\rm cos}{\phi\over2}\ket{\uparrow}
                 +{\rm sin}{\phi\over2}\ket{\downarrow},
\label{input}
\end{equation}
where the {\it a priori} density of $\phi$ is uniform,
$P(f)=P(\phi)=(2\pi)^{-1}$.  We are now to decide which template is
closest to the given $\ket{f(\phi)}$ in the sense of the highest
state overlap. We suppose further that we are given $N$ identical
copies of the input feature state $\ket{F(\phi )} = \ket{f(\phi
)}^{\otimes N}$ and the average score can be written as
\begin{equation}
\bar S(N)=\sum_{j=0}^1{1\over{2\pi}}\int_0^{2\pi} d\phi
                 \tr{\left(\hat\Pi_j\hat F(\phi)\right)}
                 \vert\amp{f(\phi)}{g_j}\vert^2,
\label {average_cost_N}
\end{equation}
where $\hat F(\phi)\equiv\proj{F(\phi)}$. Note that our score $S(C_j
| f)$ is still just $\vert\amp{f(\phi)}{g_j}\vert^2$, the overlap for
a single copy (i.e. we are establishing a relation between the input
pattern $\ket{f(\phi )}$ and the templates $\ket{g_j}$) but our POM
$\Pi_j$ operates on the full space of $N$ copies. The full input
system is described on the $N+1$ dimensional {\it totally symmetric
bosonic subspace} of ${\cal H}^{\otimes N}$, ${\cal H}_B$
\cite{Derka98,Werner98}, as
\begin{equation}
\ket{F(\phi)}\equiv\ket {f(\phi)}^{\otimes N}
          =\sum_{k=0}^N
            \sqrt{\left(\begin{array}{c} N \cr k \end{array}\right)}
            \left({\rm cos}{\phi\over2}\right)^{N-k}
            \left({\rm sin}{\phi\over2}\right)^{k}
            \ket k,
\label {ket F}
\end{equation}
where $\{\ket k\}$ is the occupation number basis of the
$\downarrow$-component. For example, in the case of $N=3$, the
basis state $\ket 2$ reads
\begin{equation}
\ket 2\equiv\left(\begin{array}{c} 3 \cr 2
\end{array}\right)^{-{1\over2}}
            \left(  \ket{\uparrow\downarrow\downarrow}
                    +\ket{\downarrow\uparrow\downarrow}
                    +\ket{\downarrow\downarrow\uparrow} \right).
\label {ket 2}
\end{equation}
Our score operators are then given by
\begin{equation}
\hat W_j\equiv
      {1\over{2\pi}}\int_0^{2\pi} d\phi \hat F(\phi)
\vert\amp{f(\phi)}{g_j}\vert^2; ~~~j=0, 1, \label
{score_op_binary}
\end{equation}
where each  $\hat W_j$ has support on the $N+1$-dimensional subspace
${\cal H}_B$ and Eq.  (\ref {average_cost_N}) can be rewritten as
\begin{equation}
\bar S(N)=\sum_{j=0}^1\tr{(\hat W_j\hat \Pi_j)}. \label
{average_cost_N_2}
\end{equation}
Without loss of generality the matching strategy $\{\hat\Pi_0,
\hat\Pi_1\}$ is constructed on ${\cal H}_B$.  In the occupation
number basis representation (using eqs. (\ref{ket F}) and
(\ref{score_op_binary})), the score operators are explicitly given as
\begin{equation}
\bra k \hat W_j \ket l=
      \sqrt{\left(\begin{array}{c} N \cr k \end{array}\right)
               \left(\begin{array}{c} N \cr l \end{array}\right)}
      {{(k+l-1)!!(2N-k-l-1)!!}\over{(2N+2)!!}}
      \left[ (N-k-l){\rm cos}{\theta_j} +(N+1)\right],
\label{Wj_even}
\end{equation}
when  $k+l$ is even, and
\begin{equation}
\bra k \hat W_j \ket l=
      \sqrt{\left(\begin{array}{c} N \cr k \end{array}\right)
               \left(\begin{array}{c} N \cr l \end{array}\right)}
      {{(k+l)!!(2N-k-l)!!}\over{(2N+2)!!}}{\rm sin}{\theta_j},
\label{Wj_odd}
\end{equation}
when $k+l$ is odd, where $\theta_0\equiv\theta$ and
$\theta_1\equiv\pi-\theta$.

\subsection{Optimal template matching}\label{Binary_temp_match_opt}

Now we consider the optimal strategy that satisfies the conditions
of Eq.  (\ref{opt_cond}).  In the present case of binary
classification, the analysis is rather straightforward, as we are
to maximize the following quantity
\begin{eqnarray}
\bar S(N)&=&\tr{(\hat W_0\hat \Pi_0)}+\tr{(\hat W_1\hat \Pi_1)} \\
            { }&=&\tr{(\hat W_1)}
                      +\tr{\left[  (\hat W_0-\hat W_1) \hat \Pi_0
\right]},
\end{eqnarray}
where the resolution of the identity $\hat \Pi_0+\hat \Pi_1=\hat
I$ was used in the second equality. Since $\tr{(\hat W_1)}=1/2$,
$\hat \Pi_0$ should be taken to maximize $\tr{\left[ (\hat
W_0-\hat W_1) \hat \Pi_0 \right]}$, that is, it should be the
projection onto the subspace corresponding to the positive
eigenvalues of the operator $\hat W_0-\hat W_1$. From Eqs.
(\ref{Wj_even}) and (\ref{Wj_odd}) we have that
\begin{equation}
\bra k(\hat W_0-\hat W_1)\ket l=
      \sqrt{\left(\begin{array}{c} N \cr k \end{array}\right)
               \left(\begin{array}{c} N \cr l \end{array}\right)}
      {{(k+l-1)!!(2N-k-l-1)!!}\over{(2N)!!}}{(N-k-l)\over (N+1)} {\rm
      cos}\theta,
\label{W0-W1}
\end{equation}
when $k+l$ is even, and $\bra k(\hat W_0-\hat W_1)\ket l=0$
otherwise. Although we have not succeeded in deriving an explicit
analytic expression for the eigenvalues $\lambda_k$ of ${\hat
W_0}-{\hat W_1}$, we introduce the diagonalizing operator $\hat P$
such that
\begin{equation}
\hat P(\hat W_0-\hat W_1)\hat P^\dagger
            =\left(\sum_{k=0}^N\lambda_k\proj{k}\right)
             {\rm cos}\theta,
\label{diag_W0-W1}
\end{equation}
where $\lambda_0>\lambda_1> .... >\lambda_N$. Since $\hat W_0 - \hat
W_1$ is antisymmetric in the antidiagonal (i.e. $(\hat W_0 - \hat W_1
)_{kl} = - (\hat W_0 - \hat W_1 )_{N-k,N-l}$) the eigenvalues match
up in $\pm$ pairs: $\lambda_N=-\lambda_0$,
$\lambda_{N-1}=-\lambda_1$, and so on (and when N is even,
$\lambda_{N/2}=0$).  The optimal strategy can then be constructed
from the pair of projection operators $\hat{\Pi}_0, \hat{\Pi}_1$ onto
the subspaces of nonnegative and negative eigenvalues respectively,
which can be written as:
\begin{mathletters}
\begin{eqnarray}
\hat\Pi_0&=\hat P^\dagger\hat\Pi_0^{\rm MV}\hat P, \quad
       \hat\Pi_0^{\rm MV}&
                 \equiv \proj{0}+\cdots +\proj{\lfloor{N\over2}\rfloor}, \\
\hat\Pi_1&=\hat P^\dagger\hat\Pi_1^{\rm MV}\hat P, \quad
       \hat\Pi_1^{\rm MV}&
                 \equiv \proj{\lfloor{N\over2}\rfloor+1} +\cdots+\proj{N},
\end{eqnarray}
\label{Pi_opt}
\end{mathletters}

\noindent where $\lfloor{N/2}\rfloor$ is the integer part of
$N/2$. The maximum average score can be finally written as
\begin{equation}
\bar S_{\rm
OPT}(N)={1\over2}+\sum_{k=0}^{\lfloor{N\over2}\rfloor}
                                   \lambda_k{\rm cos}\theta.
\end{equation}

The expressions (\ref{Pi_opt}) for the POM also provide an
intuitively appealing interpretation of the matching strategy,
which consists of two steps. The first step is the unitary
operation $\hat P$ which is applied to the $N$-product input state
$\ket{F(\phi)}$.  The second step is the measurement of the
transformed state $\hat P \hat F(\phi) \hat P^\dagger$ by the POM
$\{\hat\Pi_0^{\rm MV},\hat\Pi_1^{\rm MV}\}$. This corresponds to a
separate measurement in the $\{\ket{\uparrow},\ket{\downarrow}\}$
basis on each input copy space, followed by majority voting on the
outcomes. In other words, the transformation $\hat P$ prepares the
optimal entangled state for the final measurement, and
$\hat\Pi_0^{\rm MV}$ ($\hat\Pi_1^{\rm MV}$) provides the
projection onto the $\uparrow$-majority (the
$\downarrow$-majority) bosonic subspace. Note that for the case of
a single copy (i.e. $N=1$) $\hat W_0 - \hat W_1$ is diagonal in
the $\{\ket{\uparrow},\ket{\downarrow}\}$ basis so that $P=I$ and
the measurement in this basis is the optimal strategy.

\begin{figure}
\centerline{\psfig{file=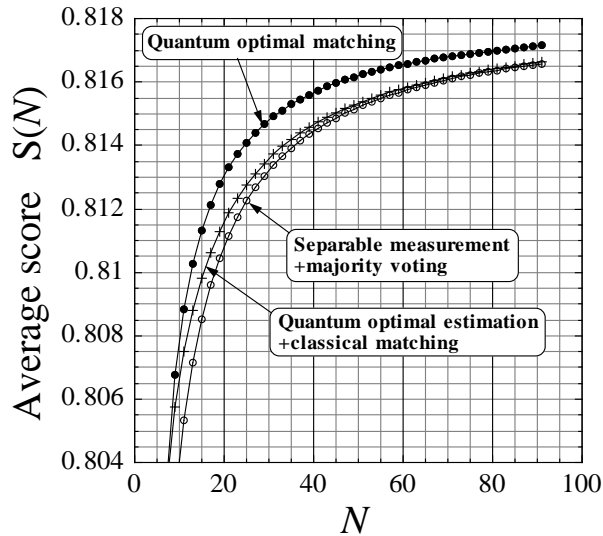,width=8cm}}
\caption{The average score in the binary template matching as a
function of the available number of copies of the input. Three
strategies are compared.  The black circles represent $\bar S_{\rm
OPT}(N)$ for the optimal strategy, while the white circles
represent $\bar S_{\rm MV}(N)$ for the strategy of separable
measurement + majority voting.  The plus correspond to $\bar
S_{\rm EST}(N, N+1, {\pi\over M})$ for the strategy of the optimal
state estimation + classical matching (section III B). }
\label{score-N_binary}
\end{figure}

In a semiclassical strategy where the separable measurement
$\{\hat\Pi_0^{\rm MV},\hat\Pi_1^{\rm MV}\}$ (corresponding to the
optimal measurement on each separate copy) is directly applied
without using the transformation $\hat P$, the attained average score
becomes, instead
\begin{equation}
\bar S_{\rm  MV}(N)={1\over2}+\sum_{k=0}^{\lfloor{N\over2}\rfloor}
        \left(\begin{array}{c} N \cr k \end{array}\right)
      {{(2k-1)!!(2N-2k-1)!!}\over{(2N)!!}}{{(N-2k)}\over{(N+1)}}
      {\rm cos}\theta.
\label{cost_MV}
\end{equation}
$\bar S_{\rm  OPT}(N)$ and $\bar S_{\rm  MV}(N)$ are compared
numerically in Fig.  \ref{score-N_binary} for the case of orthogonal
templates ($\theta =0$). The effect of $\hat P$ can be seen to reduce
the required number of sample copies to attain a prescribed level of
the average score (the curve denoted by "+" corresponds to the
strategy consisting of quantum state estimation and classical
matching, which will be explained in the next subsection).

\begin{figure}
\centerline{\psfig{file=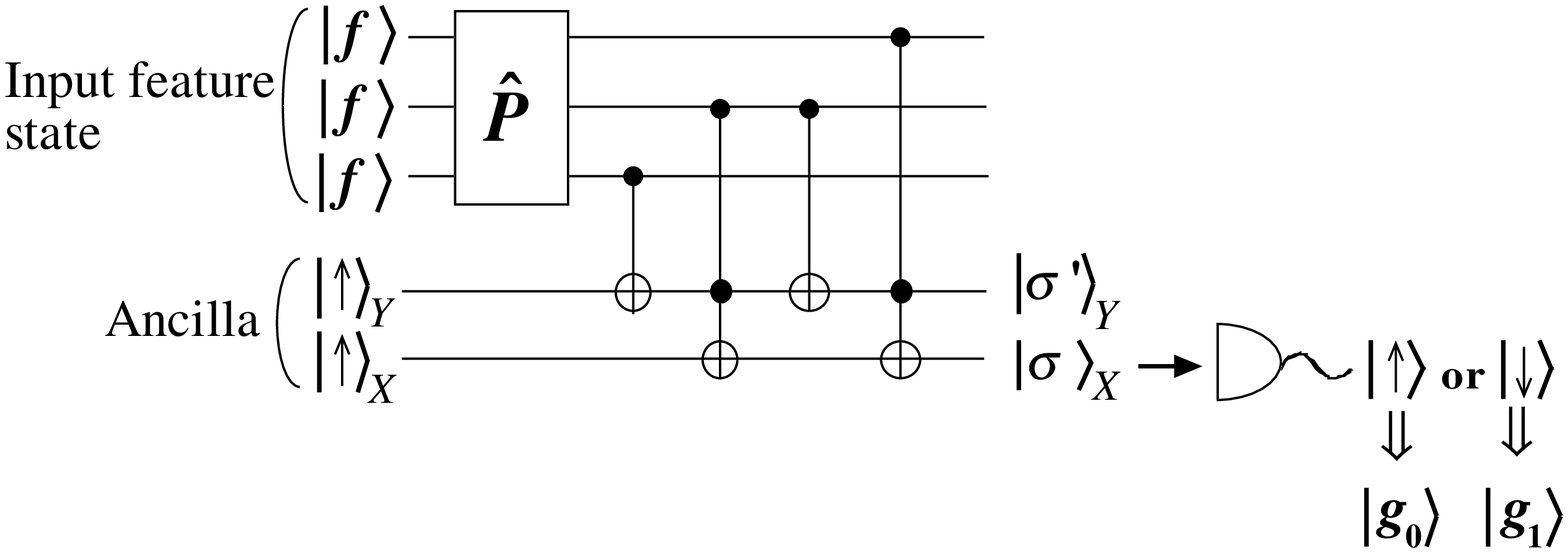,width=13cm}} \caption{A
circuit realization of the optimal classifier in the case of three
input samples. The three input samples are first transformed by
$\hat P$, and then are processed interactively with two ancillary
qubits via two controlled-NOT and two controlled-controlled-NOT
gates. In our notation $\oplus$ is the operation which
interchanges $\ket{\uparrow}$ and $\ket{\downarrow}$ and the
filled circles are control lines, i.e. the operation $\oplus$ is
applied iff all control lines are in state $\ket{\downarrow}$.
Finally, the second ancillary qubit $\ket{\sigma}_X$ is measured
in the basis \{$\ket\uparrow$, $\ket\downarrow$\}, and the two
possible outcomes imply respectively that the best matched
template is $\ket{g_0}$ or $\ket{g_1}$. }
\label{circuit_bin_temp_match}
\end{figure}

Let us illustrate the optimal matching strategy in the case where
we use three sample copies. The operator to be diagonalized is
\begin{equation}
\hat W_0-\hat W_1={{{\rm cos}\theta}\over{2^6}}
               \left(\begin{array}{cccc}
                          15&0&\sqrt {3}&0\cr
                          0&3&0&-\sqrt {3}\cr
                          \sqrt {3}&0&-3&0\cr
                          0&-\sqrt {3}&0&-15
               \end{array}\right).
\label{W0-W1_N=3}
\end{equation}
The diagonalizing matrix is found as
\begin{equation}
\hat P=    \left(\begin{array}{cccc}
                          {\rm cos}\gamma&0&{\rm sin}\gamma&0\cr
                          0&{\rm cos}\gamma&0&-{\rm sin}\gamma\cr
                          -{\rm sin}\gamma&0&{\rm cos}\gamma&0\cr
                          0&{\rm sin}\gamma&0&{\rm cos}\gamma
               \end{array}\right),
\label{P_N=3}
\end{equation}
where $\cos\gamma=[(2\sqrt{21}+9)/4\sqrt{21}]^{1/2}$ and
$\sin\gamma=[(2\sqrt{21}-9)/4\sqrt{21}]^{1/2}$. We then have
\begin{equation}
\hat P(\hat W_0-\hat W_1)\hat P^\dagger
        ={{{\rm cos}\theta}\over{2^6}}
               \left(\begin{array}{cccc}
                          {\sqrt{21}}+3&0&0&0\cr
                          0&{\sqrt{21}}-3&0&0\cr
                          0&0&-{\sqrt{21}}+3&0\cr
                          0&0&0&-{\sqrt{21}}-3
               \end{array}\right),
\label{diag_W0-W1_N=3}
\end{equation}
and
\begin{equation}
\bar S_{\rm  OPT}(3)={1\over2}
                                 +{\sqrt{21}\over 2^4}{\rm cos}\theta.
\end{equation}
One possible circuit structure for the optimal classifier is shown
in Fig.  \ref{circuit_bin_temp_match}. The input state
$\ket{F(\phi)}$
is first transformed by $\hat P$, and is then processed
interactively with two ancillary qubits via two controlled-NOT and
two controlled-controlled-NOT gates. These steps implement
$\{\hat\Pi_0^{\rm MV},\hat\Pi_1^{\rm MV}\}$ as a measurement on a
single qubit. By measuring the second ancillary qubit
$\ket{\sigma}_X$ in the basis $\{\ket\uparrow,\ket\downarrow\}$,
we can decide with the maximum average score that the best matched
template is $\ket{g_0}$ (respectively $\ket{g_1}$) when the output
is $\ket{\uparrow}$ (respectively $\ket{\downarrow}$).

\subsection{Template matching by state estimation}
\label{Binary_temp_match_estim}

Another possible kind of semiclassical strategy based on the
optimal state estimation of a qubit is also considered in Fig.
\ref{score-N_binary}. Quantum state estimation deals with how to
evaluate unknown parameters of a quantum state as precisely as
possible. This idea can be naturally applied to template matching:
one can first perform a quantum state estimation to estimate the
input feature state, and then compare this reconstructed state
with the templates {\it classically}. Optimal state estimation of
a qubit using $N$ identically prepared states has been recently
studied in Refs. \cite{Massar95,Derka98,Latorre98}. In particular,
discrete and finite element optimal POMs were found
\cite{Derka98,Latorre98}, and they maximize the following score
\begin{equation}
\bar S(N)\equiv
  \sum_m{1\over{2\pi}}\int_0^{2\pi} d\phi
   \tr{ \left(  \hat\Pi_m  \hat f(\phi)^{\otimes N}  \right) }
   \vert\langle f(\phi)\vert f(\phi_m) \rangle \vert^2.
\label{score_N_phi}
\end{equation}
where $\vert f(\phi_m)\rangle$ is a reconstructed state after the
state estimation. The assignment of a guessed state
$m\rightarrow\vert f(\phi_m)\rangle$ is now also to be optimized.
This strategy was already described in Ref. \cite{Derka98}, but we
rephrase it here in a slightly different and more practical way
according to the results of Ref. \cite{Sasaki_Carlini2000b}. For
convenience of calculation we introduce a new basis $\{
\ket{v_0},\ket{v_1} \}$ for which our great circle of feature
states $\ket{f(\phi )}$ is the equator. We fix the basis vectors
by requiring the templates to have the symmetrical form:
\begin{mathletters}
\begin{eqnarray}
\ket{g_0}&=&{1\over\sqrt{2}}
             \left(  {\rm e}^{ -i( {\pi\over4} - {\theta\over2} ) }
                         \ket{v_0}
                      +{\rm e}^{   i( {\pi\over4} - {\theta\over2} ) }
                         \ket{v_1}
             \right), \\
\ket{g_1}&=&{1\over\sqrt{2}}
             \left(  {\rm e}^{   i( {\pi\over4} - {\theta\over2} ) }
                         \ket{v_0}
                      +{\rm e}^{ -i( {\pi\over4} - {\theta\over2} ) }
                        \ket{v_1}\right).
\label{templates_v0v1}
\end{eqnarray}
\end{mathletters}
and the circle of feature states may be taken to be
\begin{equation}
\ket{f(\phi)}\equiv {1\over\sqrt{2}} \left( {\rm
e}^{-i{\phi\over2}}\ket{v_0}
         +{\rm e}^{ i{\phi\over2}}\ket{v_1} \right).
\label{input_v0v1}
\end{equation}
where again, the parameter $\phi$ is uniformly distributed over
$[0,2\pi )$.

Let us also introduce $M$ states equally spaced on the Bloch great
circle (which will define our POM):
\begin{equation}
\ket{f_m(\varphi)}={1\over\sqrt{2}}
             \left [  {\rm e}^{-i( {\varphi\over2}
                                          +{{m\pi}\over M})}\ket{v_0}
                       +{\rm e}^{  i( {\varphi\over2}
                                          +{{m\pi}\over
M})}\ket{v_1}\right]; \quad (m=0, 1, ..., M-1)
\label{uniform_templates_varphi}
\end{equation}
Here we have introduced a phase factor $\varphi$
which determines the position of these symmetrical
states relative to the fixed positions of the template states. The
corresponding $N$-fold tensor product states are:
\begin{equation}
\ket{F_m(\varphi)}\equiv\ket{f_m(\varphi)}^{\otimes N}
          =\sum_{k=0}^N
            \sqrt{{1\over2^N}
            \left(\begin{array}{c} N \cr k \end{array}\right)}
            {\rm e}^{ -i(N-2k) ( {\varphi\over2}+{{m\pi}\over M} ) }
            \ket{k}_v,
\label{ket F tilde}
\end{equation}
where $\{\ket{k}_v\}$ is the symmetric bosonic basis for
$\{\ket{v_0},\ket{v_1}\}$. It can then be shown that the {\it square
root measurement} $\{\ket{\mu_m}\bra{\mu_m}\}$ based on the states
$\{\ket{F_m(\varphi)}\}$, that is,
\begin{equation}
\ket{\mu_m(\varphi)} \equiv \left(
                \sum_{m=0}^{M-1}
                \ket{F_m(\varphi)}\bra{F_m(\varphi)}
            \right)^{-{1\over2}}
            \ket{F_m(\varphi)},
\label{srm}
\end{equation}
and the associated guess $\{\ket{f_m(\varphi)}\}$, provides an
optimal state estimation strategy when we take $M>N$
\cite{Sasaki_Carlini2000b}.

Thus, by applying the POM
$\{\vert\mu_m(\varphi)\rangle\langle\mu_m(\varphi)\vert\}$, the input
feature state is optimally reconstructed as one of the
$\ket{f_m(\varphi)}$. Then one can {\it classically} compare this
reconstructed state with the templates and pick up the template state
which has the largest overlap with the reconstructed state. The above
strategy for state {\em estimation} is optimal for any choice of
$\varphi$. Indeed since our input state distribution is uniform, the
score in eq. (\ref{score_N_phi}) will be independent of $\varphi$.
But when we compare the reconstructed state with the templates, the
resulting average score of template matching {\em will} depend on
$\varphi$ due to the fixed positions of the templates, i.e. different
state estimation strategies which each give the same best possible
score will generally give different scores for template matching via
our classical method, and we should choose the best from our set of
optimal estimation strategies. To complete our semi-classical
template matching procedure we categorize the  $\ket{f_m(\varphi)}$'s
into two classes according to the template states, that is, condense
the set $\{\vert\mu_m(\varphi)\rangle\langle\mu_m(\varphi)\vert\}$
into a two element POM $\{\hat\Pi_0^{\rm EST}(\varphi),\hat\Pi_1^{\rm
EST}(\varphi)\}$, whose elements indicate that the best matched
template is $\ket{g_0}$ or $\ket{g_1}$, respectively.

For simplicity let us assume that $M$ is even.
Then by symmetry, it is enough to consider $\varphi$ in the
range $[0, {{2\pi}\over M})$.
The categorization boundary is determined by the condition
\begin{equation}
 \vert\langle g_0\vert f_m(\varphi)\rangle\vert
=\vert\langle g_1\vert f_m(\varphi)\rangle\vert
\end{equation}
The values of $m$ with $ \vert\langle g_0\vert
f_m(\varphi)\rangle\vert
  \ge \vert\langle g_1\vert f_m(\varphi)\rangle\vert $
are categorized into the class of $\ket{g_0}$, and the others into
that of $\ket{g_1}$. Noting that
\begin{equation}
 \vert\langle g_0\vert f_m(\varphi)\rangle\vert^2
-\vert\langle g_1\vert f_m(\varphi)\rangle\vert^2 =2{\rm sin}
(\varphi+{{2m\pi}\over M})
    {\rm sin} ({\pi\over2}-\theta),
\end{equation}
the binary categorization should then be
\begin{mathletters}
\begin{eqnarray}
\hat\Pi_0^{\rm EST}(\varphi)&=&\sum_{m=0}^{{M-2}\over2}
\vert\mu_m(\varphi)\rangle\langle\mu_m(\varphi)\vert, \\
\hat\Pi_1^{\rm EST}(\varphi)&=&\sum_{m={M\over2}}^{M-1}
\vert\mu_m(\varphi)\rangle\langle\mu_m(\varphi)\vert.
\label{Pi_EST}
\end{eqnarray}
\end{mathletters}
The average score for this strategy is
\begin{mathletters}
\begin{eqnarray}
\bar S_{\rm EST}(N, M, \varphi) &=&
\sum_{j=0}^1{1\over{2\pi}} \int d \phi
      \tr{\left[ \hat\Pi_j^{\rm EST}(\varphi)
                      \hat f(\phi)^{\otimes N}
            \right]}
      \vert\amp{f(\phi)}{g_j}\vert^2\\
{ }&=&\tr{\left[\hat W_0\hat\Pi_0^{\rm EST}(\varphi)\right]}
          +\tr{\left[\hat W_1\hat\Pi_1^{\rm EST}(\varphi)\right]},
\label{average_cost_N_EST}
\end{eqnarray}
\end{mathletters}
\noindent where
\begin{equation}
\hat W_0={1\over2^{N+2}}
     \left[
              2\sum_{k=0}^N
                 \left(\begin{array}{c} N \cr k \end{array}\right)
                 \vert k\rangle_v {}_v\langle k\vert
           + \sum_{k=0}^{N-1}
                 \left(\begin{array}{c} N \cr k \end{array}\right)
                 \sqrt{{N-k}\over{k+1}}
                 \left(
                           {\rm e}^{ i( {\pi\over2}-\theta ) }
                            \vert k+1\rangle_v {}_v\langle k\vert
                         +{\rm e}^{-i( {\pi\over2}-\theta ) }
                            \vert k\rangle_v {}_v\langle k+1\vert
                 \right)
     \right]=\hat W_1^\dagger.
\label{W0_varphi}
\end{equation}
By a straightforward calculation we obtain
\begin{mathletters}
\begin{eqnarray}
\bar S_{\rm EST}(N, M, \varphi)&=&{1\over2}+
              { {{\rm cos}\theta{\rm cos} (\varphi-{\pi\over M})}
                 \over
                 {2^N M{\rm sin} ({\pi\over M})}  }
                \sum_{k=0}^{N-1}
                       \left(\begin{array}{c} N \cr k
\end{array}\right)
                       \sqrt{{N-k}\over{k+1}} \\
{ }&<&\bar S_{\rm EST}(N, N+1, {\pi\over M}) \\ { }&=&
              {1\over2}+
              { {{\rm cos}\theta}
                 \over
                 {2^N (N+1){\rm sin} ({\pi\over{N+1}})} } \sum_{k=0}^{N-1}
                 \left(\begin{array}{c} N \cr k \end{array}\right)
                 \sqrt{{N-k}\over{k+1}}.
\label{average_cost_N_EST_2}
\end{eqnarray}
\end{mathletters}

\noindent The quantity $\bar S_{\rm EST}(N, N+1, {\pi\over M})$ is
compared with the optimal score $\bar S_{\rm OPT}(N)$ and the one
$\bar S_{\rm MV}(N)$ obtained by the separable measurement plus
majority voting scheme in Fig. \ref{score-N_binary} (for
$\theta=0$). As it can be seen, the strategy using the optimal
state estimation followed by classical matching can be close to
optimal for the region of small $N$, while as $N$ increases it
starts to deviate from the optimal one and becomes closer to the
strategy of separable measurement plus majority voting.

\begin{figure}
\centerline{\psfig{file=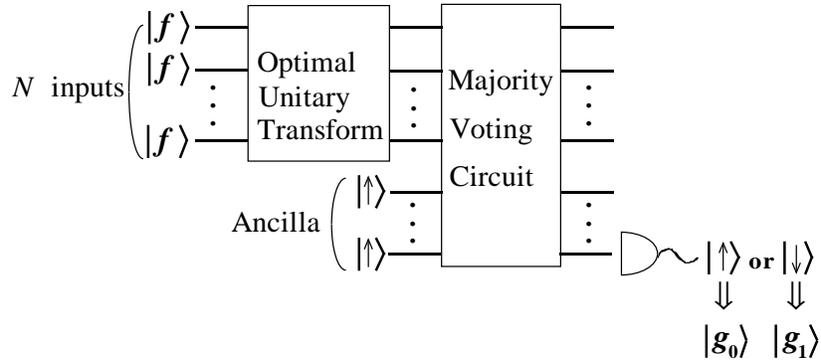,width=11cm}} \caption{A
scheme of the optimal classifier for $N$ input samples which is a
generalization of Fig. \protect\ref{circuit_bin_temp_match}. The
binary classification of interest would eventually be turned into
the measurement of a single qubit in the basis
$\{\vert\uparrow\rangle,\vert\downarrow\rangle\}$. }
\label{circuit_OPT}
\end{figure}

\begin{figure}
\centerline{\psfig{file=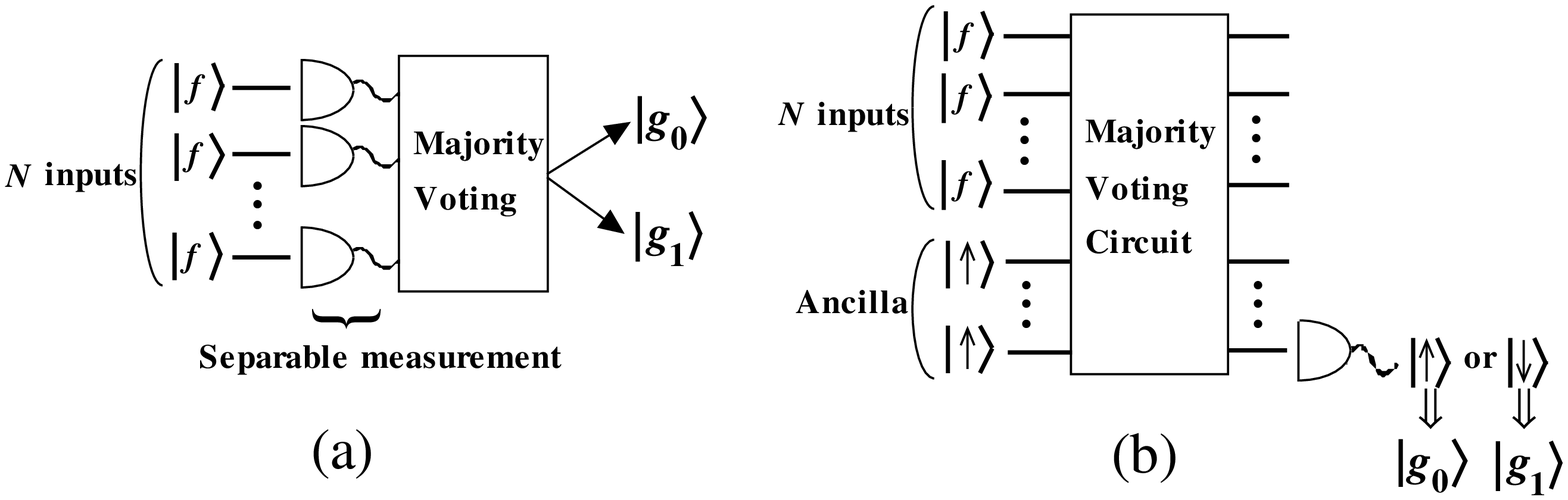,width=18cm}}
\caption{Schemes of the strategy for the separable measurement
plus majority voting.  (a) is the direct translation of the POM
which includes $N$ measurements.  But this can be translated into
a measurement on a single ancillary qubit plus an additional
circuit (majority voting circuit) beforehand as shown in (b). The
majority voting circuit includes a series of C-NOT gates just as
in Fig. \protect\ref{circuit_bin_temp_match}.  }
\label{circuit_MV}
\end{figure}

\begin{figure}
\centerline{\psfig{file=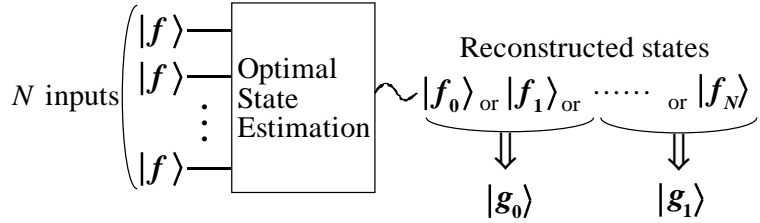,width=10cm}} \caption{ A
scheme of the strategy for the optimal state estimation plus
classical matching.  The optimal state estimation is a collective
measurement on $N$ identical copies of the sample state.  By
applying this, the input feature state is reconstructed.  The
output would be one of the $N+1$ candidates $\ket{f_m}$.  Then one
can compare this reconstructed state with the templates {\it
classically}.  This is actually a categorization of $\ket{f_m}$
into two classes according to the template states.}
\label{circuit_ETS}
\end{figure}

The three strategies are schematically summarized in Figs.
\ref{circuit_OPT}$\sim$\ref{circuit_ETS}. The quantum optimal
strategy (Fig. \ref{circuit_OPT}) is realized by a collective
measurement on the state ${\ket f}^{\otimes N}$ with binary
outputs. This is made by dividing the state space ${\cal H}_B$
spanned by ${\ket f}^{\otimes N}$ into 2 parts according to the
templates, and by successfully using entanglement effects in
${\cal H}_B$. On the other hand, the separable measurement plus
majority voting scheme shown in Fig. \ref{circuit_MV} does not
take any advantage of the entanglement which could be drawn from
the state ${\ket f}^{\otimes N}$. In the optimal state estimation
plus classical matching strategy shown in Fig. \ref{circuit_ETS},
the collective measurement first performed for estimating ${\ket
f}$ also utilizes an entanglement effect. However, this is not the
best way for {\it binary classification}. In fact, the optimal
state estimation requires dividing the space ${\cal H}_B$ into at
least $N+1$ parts. As $N$ increases, one has to rely much more on
the classical procedure to categorize the outputs into two
classes. This is the reason why this strategy becomes ineffective
for larger $N$. Intuitively any intermediate measurement prior to
the final decision tends to degrade the total performance leading
to a waste of input copies for a given average score level, so the
process for the best binary classification should stay entirely in
the quantum domain until the very final measurement.


\section{Multiple template matching of a two state
system}\label{Multiple_temp_match}

In the previous section we have assumed a single feature parameter
$\phi$ and a minimum number (two) of templates. In this section we
extend our model to allow for multiple template matching. Although
binary template matching can be reduced to the diagonalization of
the operator $\hat W_0-\hat W_1$, there is no such straightforward
method to find the optimal strategy in general cases. To keep the
model tractable, we assume that the input feature state is a
general qubit state depending now on two parameters,
\begin{equation}
\ket{f} ={\rm e}^{-i{\phi\over2}}{\rm
cos}{\theta\over2}\ket{\uparrow} +{\rm e}^{ i{\phi\over2}}{\rm
sin}{\theta\over2}\ket{\downarrow}. \label{input_general}
\end{equation}
with a uniform {\em a priori} distribution over the whole Bloch
sphere. Furthermore we suppose that only one of the parameters
relates to the desired feature of $\ket{f}$, for example, the angle
parameter $\phi$ around the $\hat\sigma_z$ axis, while the
$\hat\sigma_z$ component itself is of no interest. The template
states corresponding to this feature are assumed to be $M$ states
uniformly distributed around the great circle in the $x-y$ plane of
the Bloch sphere, that is,
\begin{equation}
\ket{g_m}={1\over\sqrt{2}}
             \left(  {\rm e}^{-i{{m\pi}\over M}}\ket{\uparrow}
                     +{\rm e}^{ i{{m\pi}\over
M}}\ket{\downarrow}\right); \quad (m=0, 1, ..., M-1).
\label{uniform_templates_g}
\end{equation}
As before we have $N$ copies of the input state as
\begin{equation}
\ket F\equiv\ket f^{\otimes N}
      =\sum_{k=0}^N
         \sqrt{\left(\begin{array}{c} N \cr k \end{array}\right)}
         \left(  {\rm e}^{-i{\phi\over2}}
                    {\rm cos}{\theta\over2}
         \right)^{N-k}
         \left(  {\rm e}^{ i{\phi\over2}}
                    {\rm sin}{\theta\over2}
         \right)^{k}
         \ket k,
\label{ket F_general}
\end{equation}
and generate the score operators based on this and the templates
as
\begin{mathletters}
\begin{eqnarray}
\hat W_m&\equiv&
   {1\over{4\pi}} \int_0^{2\pi} d\phi
                          \int_0^\pi d\theta {\rm sin}\theta
   {\ket F}{\bra F} ~\vert\amp{f}{g_m}\vert^2 \\
{ }&=& {1\over2(N+1)}
          \left[
             \hat I
           +\sum_{k=0}^{N -1}
              {\sqrt{(N-k)(k+1)}\over{N+2}}
              \left(  {\rm e}^{-i{{m\pi}\over M}}
                          \vert k+1\rangle\langle k\vert
                       +{\rm e}^{ i{{m\pi}\over M}}
                          \vert k\rangle\langle k+1\vert
              \right)
          \right]  ,
\label{Wm_general}
\end{eqnarray}
\end{mathletters}

\noindent We then seek the strategy to find the template which
best matches with the given $\ket{f}$ in such a way  to maximize
the average score
\begin{equation}
\bar S=\sum_{m=0}^{M-1}\tr{(\hat W_m\hat \Pi_m)}.
\label{average_score_general}
\end{equation}
As noted in section \ref{Bayesian} this optimal template problem
is equivalent to the problem of optimal discrimination of the set
of mixed states $\frac{1}{2}\hat{W}_m$ taken with equal {\em a
priori} probabilities $p_m = 1/M$.

The score operators evidently have the same symmetry as the
templates, that is,
\begin{equation}
\hat W_m= \hat V^m \hat W_0\hat V^{\dagger m},
\label{Wm_general_symmetry}
\end{equation}
where
\begin{equation}
\hat V\equiv\sum_{k=0}^N{\rm e}^{-i{(N-2k)\pi\over
M}}\ket{k}\bra{k},
\end{equation}
is a unitary representation of the group of integers modulo $M$ on
the $N+1$ dimensional bosonic subspace of $N$ qubits. Indeed it is
just the product representation $\hat V = \hat{v}^{\otimes N}$ where
$\hat v$ is the operation of rotation of the one-qubit Bloch sphere
by $2\pi/M$ about the $z$-axis. Now it is known
\cite{Holevo73_condition} that for any group covariant set of states,
the state discrimination problem always has an optimal strategy that
is similarly group covariant, i.e. there will be an optimal POM of the
form $\hat\Pi_m=\hat V^m\hat\Pi_0\hat V^{\dagger m}$ and the
optimality conditions reduce to \cite{Holevo_book}
\begin{equation}
\begin{array}{cl}
\mbox{(i')}&\hat\Gamma\equiv\displaystyle
    \sum_{m=0}^{M-1} \hat V^m\hat W_0\hat \Pi_0\hat V^{\dagger m} \mbox{ is
hermitian,}
\\
\mbox{(ii')}&\hat\Gamma-\hat W_0\ge0.
\end{array}
\label{reduced opt_cond}
\end{equation}

We have succeeded in deriving an optimal strategy only in the case
that $M>N$, i.e. when the number of copies is less than the number
of templates.  This is again the square root measurement built
from the templates $\ket{G_m}\equiv\ket{g_m}^{\otimes N}$, that
is, the set \{$\hat\Pi_m=\ket{\mu_m}\bra{\mu_m}$\} with
\begin{equation}
\ket{\mu_m}\equiv \hat G^{-{1\over2}}\ket{G_m}, \quad \hat
G=\sum_{m=0}^{M-1}\ket{G_m}\bra{G_m}.
\end{equation}
In fact, by using the orthogonality relation
\begin{equation}
\sum_{m=0}^{M-1}{\rm exp}(i{{2m\pi}\over M}n)=M\delta_{n,0}
\quad\mbox{ for }-M< n< M, \label{orthogonality}
\end{equation}
(so $\hat G$ like $\hat V$ is diagonal in the $\ket{k}$ basis and
$\hat{G}^{1/2}$ commutes with $\hat V$), we find that
\begin{equation}
\ket{\mu_m}=\hat V^m \ket{\mu_0}, \quad
\ket{\mu_0}={1\over\sqrt M}\sum_{k=0}^N\ket{k}.
\label{ket mu}
\end{equation}
The optimality of this POM can then be proved by checking the
conditions (i') and (ii') directly as follows. From Eq.
(\ref{orthogonality}), we obtain
\begin{equation}
\hat\Gamma={1\over2(N+1)}
          \left[\hat I
           +\sum_{k=0}^{N -1}
              {\sqrt{(N-k)(k+1)}\over{N+2}}
              \left(  \vert k\rangle\langle k\vert
                     +\vert k+1\rangle\langle k+1\vert
              \right)
         \right],
\label{Gamma}
\end{equation}
and, consequently,
\begin{equation}
\hat\Gamma-\hat W_0={1\over2(N+1)(N+2)}
           \sum_{k=0}^{N -1}
              \sqrt{(N-k)(k+1)}
          \Big[\vert k\rangle\langle k\vert
                        +  \vert k+1\rangle\langle k+1\vert
                      -\vert k\rangle\langle k+1\vert
                        - \vert k+1\rangle\langle k\vert \Big].
\label{Gamma-W0}
\end{equation}
Since each $2\times2$ matrix inside the brackets $[...]$ in Eq.
(\ref{Gamma-W0}) has the eigenvalues 0 and 2 and is non-negative
definite, so also is $\hat\Gamma-\hat W_0$ and (ii') of Eq.
(\ref{reduced opt_cond}) holds. Condition (i') of Eq. (\ref{reduced
opt_cond}) can be checked in a straightforward manner from Eqs.
(\ref{ket mu}) and (\ref{Gamma-W0}). The maximum average score does
not depend on $M$ and reads
\begin{equation}
\bar S_{\rm max}(N)=M\tr{(\hat W_0\hat \Pi_0)}
              ={1\over2}+\sum_{k=0}^{N -1}
                {\sqrt{(N-k)(k+1)}\over{(N+1)(N+2)}}.
\end{equation}
We also note that $\ket{G_0}$ is the maximum-eigenvalue eigenstate
of $\hat W_0$, i.e. the spectral decomposition is
\begin{equation}
\hat W_0=\sum_{k=0}^N\omega_k\proj{\omega_k}, \quad
\omega_k={{k+1}\over(N+1)(N+2)},
\label{W_0 spectral decomp}
\end{equation}
with $\ket{G_0}=\ket{\omega_N}$. This is especially interesting in
view of the following theorem proved in \cite{Holevo73_condition}:
\\ Theorem: Let $G$ be a group and let $g\rightarrow \hat{V}_g$ be
an {\em irreducible} representation of $G$ on a $d$ dimensional
Hilbert space $\cal H$. Let $\{ \hat{F}_g : g\in G \}$ be a
collection of Hermitian operators on $\cal H$ such that $\hat{F}_g
= \hat{V}_g \hat{F}_e\hat{V}_g^\dagger$ (where $e$ is the identity
of $G$). For any POM $X= \{ \hat{X}_g : g\in G \}$ consider the
function
\[ Q(X)= \tr \sum_g \hat{F}_g \hat{X}_g . \] Let $Z =
\frac{d}{|G|} \proj{\phi}$ where $\ket{\phi}$ is the maximum
eigenvalue eigenstate of $\hat{F}_e$ (and $|G|$ is the size of
$G$).\\ Then $Q$ is maximized by the POM $\{ \hat{V}_g Z
\hat{V}_g^\dagger : g\in G \}$.

Note that $\hat{V}_g \ket{\phi}$ is a maximum eigenvalue eigenstate
of $\hat{F}_g$ so the theorem claims that the $G$ covariant POM based
on these $|G|$ eigen-directions is optimal. By irreducibility of the
representation we have (via Schur's lemma) that $\sum_g \hat{V}_g
\hat{A} \hat{V}_g^\dagger$ is a multiple of the identity for any
operator $\hat{A}$. Thus the square root measurement construction
does not alter these maximal eigen-directions when the representation
is irreducible. In our template matching problem $G$ is the group of
integers modulo $M$ and $\hat{F}_g$ correspond to the score operators
$\hat{W}_m$. $\cal H$ is the $N+1$ dimensional bosonic subspace of
$N$ qubits and the group acts via $m\rightarrow \hat{V}^m$. This
representation is {\em not} irreducible so the theorem does not
apply. Yet we have shown that an optimal measurement is still
obtainable from the maximum eigenvalue eigenstates of the score
operators. In this case (of a reducible representation) the square
root construction will give a non-trivial change in the directions of
the maximal eigenstates, necessary to obtain a POM from them. This
suggests a possible avenue of generalization for the above theorem of
\cite{Holevo73_condition} which we will explore elsewhere.


In the other case $M\le N$, that is, when we can use a larger
number of copies of the input than the number of templates, the
optimal POM is more complicated. This should include elements with
rank 2 or higher because of the requirement that
$\sum_{m=0}^{M-1}\hat\Pi_m=\hat I$ in the $N+1$ dimensional
bosonic subspace. We have not yet found a systematic way to
construct such higher rank POMs. Here we discuss some simple
cases.

The simplest case is $M=2$, that is, binary classification. In
this case, the two score operators commute and the strategy of
separable measurement in the binary template basis on each copy
plus majority voting turns out to be optimal (note that the
binary template problem in section \ref{Binary_temp_match} had a
different distribution of input states and the the two templates
there were not required to be orthogonal).

The next simplest case is $M=N=3$. The optimal POM is specified by
\begin{equation}
\hat\Pi_0=\left(\begin{array}{cccc}
          {1\over3}&a&c&0 \cr
          a&{1\over3}&b&c \cr
          c&b&{1\over3}&a \cr
          0&c&a&{1\over3}
          \end{array}\right),
\end{equation}
with $a=(\sqrt{21}+\sqrt{5})/24$, $c=(\sqrt{35}-\sqrt{3})/24$, and
$b=6ac$, and the maximum average score is
\begin{equation}
\bar S_{\rm max}(N) =\tr{\hat\Gamma}={{5+3\sqrt{3}a+3b}\over10}.
\end{equation}
This $\hat \Pi_0$ is derived by solving the equations for the
condition (\ref{reduced opt_cond})-(i') directly and then by
picking up the solution satisfying the condition (\ref{reduced
opt_cond})-(ii'). $\hat \Pi_0$ is a rank 2 operator
\begin{equation}
\hat\Pi_0=\lambda_+\proj{\lambda_+}+\lambda_-\proj{\lambda_-},
\end{equation}
with $\lambda_+=0.964$ and $\lambda_-=0.370$ and
\begin{mathletters}
\begin{eqnarray}
\ket{\lambda_+}&=&0.995\ket{\omega_3}+0.100\ket{\omega_1},\\
\ket{\lambda_-}&=&0.979\ket{\omega_2}+0.204\ket{\omega_0},
\label{lambda}
\end{eqnarray}
\end{mathletters}

\noindent
where
$\ket{\omega_0},\ket{\omega_1}, \ket{\omega_2}, \ket{\omega_3}$
are the eigenstates of $\hat W_0$ corresponding to eigenvalues in
increasing order (Eq. (\ref{W_0 spectral decomp})).
Thus, although the main component of $\hat\Pi_0$
comes from the maximum-eigenvalue eigenstate $\ket{\omega_3}$ of
$\hat W_0$, the other eigenstates are also involved with
appropriate weights.

Finally we mention the case of $M=3$ and $N=4$. The optimal POM is
specified by
\begin{equation}
\hat\Pi_0=\left(\begin{array}{rrrrr}
          {1\over3}&a&c&0&-\sqrt{3\over8}c \cr
          a&{1\over3}&b&\sqrt{3\over8}c&0 \cr
          c&b&{1\over3}&b&c \cr
          0&\sqrt{3\over8}c&b&{1\over3}&a  \cr
          -\sqrt{3\over8}c&0&c&a&{1\over3}
          \end{array}\right),
\end{equation}
with $b=\sqrt{29+\sqrt{201}}/24$,
$c=(\sqrt{67}-\sqrt{3})/(24\sqrt{2})$, and $a=6bc$. The structure
of $\hat \Pi_0$ is again of the form
\begin{equation}
\hat\Pi_0=\lambda_+\proj{\lambda_+}+\lambda_-\proj{\lambda_-},
\end{equation}
with $\lambda_+=1$ and $\lambda_-=2/3$,
\begin{mathletters}
\begin{eqnarray}
\ket{\lambda_+}&=&0.992\ket{\omega_4}+0.115\ket{\omega_2}
                               +0.044\ket{\omega_0},\\
\ket{\lambda_-}&=&0.984\ket{\omega_3}+0.178\ket{\omega_1},
\end{eqnarray}
\end{mathletters}

\noindent and the maximum average score is
\begin{equation}
\bar S_{\rm max}(N) ={{5+4a+2\sqrt{6}b}\over10}.
\end{equation}


Generally speaking it is more difficult to find analytic solutions
for the Bayes optimal strategy for mixed states, and one has to
rely on numerical methods. The above examples indicate that the
largest eigenvalue eigenstates of the score operator $\hat W_0$
should play an essential role in constructing the optimal POM (and
maximizing the score), while smaller eigenvalue eigenstates can be
regarded as perturbative correction terms. This might be helpful
for considering efficient numerical algorithms for finding the
optimal POM.

\section{Concluding remarks}\label{Concluding}

We have considered the problem of quantum template matching, which is
to find the template state that best matches a given input feature
state. The quality of matching was taken to be the standard overlap
of quantum states. This question was formulated in the context of
quantum Bayesian inference and it was seen to be equivalent to the
optimal discrimination of certain mixed states given in terms of
score operators, each defined for a specific template state and
including all the {\it a priori} information about the input.

In this paper, the simplest case of binary classification of a two
state system with a single feature parameter was extensively
studied. We constructed the optimal strategy in the $N+1$
dimensional bosonic state space ${\cal H}_B$ spanned by the tensor
products $\ket{f}^{\otimes N}$ of $N$ identical copies of the
input state. The optimal state estimation on $\ket{f}^{\otimes N}$
followed by a classical matching process does not provide the best
strategy, and there is a different optimal use of entanglement for
this particular binary classification problem. In the case of
multiple template matching, the problem becomes more difficult and
we derived the optimal strategies in a few illustrative cases.

As mentioned in the introduction, the procedure of conventional
pattern matching consists of feature extraction, calculation of
the discriminant function and classification. In the quantum
context, however, it is not clear how to model such processes
without an associated loss of useful information. For instance, to
eliminate features of no concern one might simply project the
input state onto the subspace spanned by the relevant states with
the features of interest. But we saw in section
\ref{Binary_temp_match} that a quantum measurement, that is, a
projection of states, carried out before the final template
decision, is generally detrimental to optimal performance. In this
spirit, we dealt with the problem in the original Hilbert space
without projecting the input states onto the subspace for the
features of interest (section \ref{Multiple_temp_match}), and the
whole process of quantum template matching was represented by a
single POM. It is of course an open question to formulate quantum
protocols in more physically comprehensive ways, e.g. involving a
separate non-trivial feature enhancement process prior to
classification, and to systematically derive optimal strategies
for them.

But even without such additional infrastructure (e.g. feature
enhancement) our problem of template matching has some interesting
generalizations related to the role of classical versus quantum
information in the formulation. In our formulation we have assumed
that the input states (such as $\ket{f}$) are given as quantum
information (i.e. unknown quantum states) whereas the template states
($\ket{g_i}$'s with known identities) are given as classical
information. Furthermore our goal was to obtain the best template as
classical information (i.e. knowledge of the identity of the best
$\ket{g_i}$) via a suitable POM. The ingredients of this formulation
can be relaxed in a variety of potentially interesting ways and here
we mention two such ways: \\ (a) Instead of knowing the identities of
the template states we may merely be given only some finite number
($K$) of copies of each template (so our original formulation is
equivalent to $K=\infty$). One matching strategy would then be to
apply state estimation to the sets of $K$ copies and proceed as in
our original formulation with the resulting estimated state
identities. But this is unlikely to be optimal and we should consider
a more fully quantum procedure which, for any input $\ket{f}$,
identifies the best template class (still here as classical
information) without attempting to obtain any further information
about the identities of the template states themselves. \\ (b) A
second more intrinsically quantum mechanical formulation of template
matching involves obtaining the answer (i.e. the best matching
template) only as quantum information. In this scenario we have a
known prior distribution of inputs $\{ \ket{f_i}; p_i \}$ and a known
set of possible templates $\{ \ket{g_j} \}$. Then given one (or more)
copies of $\ket{f_i}$ we want to design a quantum process (i.e. a
completely positive trace preserving map acting on the input) that
outputs (one copy of) a quantum state $\sigma_i$ of the form
$\ket{f_i} \rightarrow \sigma_i = \sum_j p_{ij} \proj{g_j}$ such that
some suitable average score $\sum_{ij}p_i  p_{ij} S(j|i)$ is
maximized. Note that the formulation in our paper (of getting the
best template as classical information) would provide one possible
strategy since we can then construct the corresponding template state
as a quantum state, but again, this would not be expected to be
optimal since we produce a great deal of unwanted extra information
in addition to the desired quantum output state.

There are yet further possible avenues for generalizing the
formulation of the template matching problem. One is to study pattern
classification with other kinds of matching criteria than fidelity,
which would be chosen according to some specific application or
purpose. For example, according to the quantum Sanov theorem (e.g.
summarized in section IV of \cite{VP}) the quantum relative entropy
$S(\hat f || \hat g )$ between two quantum states provides an index
for estimating the probability that the states will not be
distinguished on the basis of an arbitrary measurement on $N$ copies
of the state. Thus the relative entropy provides an alternative,
operationally intuitive, notion of ``distance'' between quantum
states and we may consider maximizing the average relative entropy as
our similarity criterion in template matching for some purposes.

The above remarks and generalizations show that the problem of
template matching introduced in this paper is just the beginning of a
fruitful area for further study. The formulation adopted in the paper
is perhaps the simplest, in that it is closely related to an existing
body of results on quantum Bayesian estimation. But a study of
possible hybrid quantum-classical generalizations along the lines
suggested above would provide a natural setting for characterizing
new properties, and a deeper understanding, of quantum information
itself, and especially the ways it fails to accord with familiar
properties of classical information.

\acknowledgements

The authors would like to thank T. Hattori for introducing them to
the pattern matching problem and for giving valuable suggestions.
They would also like to thank V. Buzek, A.  S. Holevo, C.  A.
Fuchs, C.  H. Bennett, O.  Hirota, S.  M.  Barnett and A.  Chefles
for helpful discussions. A.C.'s research is supported by JISTEC
under grant no. 199016. R.J. is supported by the U.K. Engineering
and Physical Sciences Research Council.

\end{document}